\documentclass[aps,prl,preprint,preprintnumbers,amsmath,amssymb,citeautoscript,floatfix,superscriptaddress,showkeys,showpacs]{revtex4-1}

\usepackage[latin1]{inputenc}
\usepackage{graphicx}

\begin{document}

\author{Markus R. Wagner}
\email{*wagner.markus.r@gmail.com}
\affiliation{Catalan Institute of Nanoscience and Nanotechnology (ICN2), CSIC and The Barcelona Institute of Science and Technology, Campus UAB, Bellaterra, 08193 Barcelona, Spain}
\affiliation{Institute of solid state physics, Technical University Berlin, Hardenbergstr. 36, 10623 Berlin, Germany}
\author{Bartlomiej Graczykowski}
\author{Juan Sebastian Reparaz}
\author{Alexandros El Sachat}
\author{Marianna Sledzinska}
\author{Francesc Alzina}
\affiliation{Catalan Institute of Nanoscience and Nanotechnology (ICN2), CSIC and The Barcelona Institute of Science and Technology, Campus UAB, Bellaterra, 08193 Barcelona, Spain}
\author{Clivia M. Sotomayor Torres}
\affiliation{Catalan Institute of Nanoscience and Nanotechnology (ICN2), CSIC and The Barcelona Institute of Science and Technology, Campus UAB, Bellaterra, 08193 Barcelona, Spain}
\affiliation{ICREA - Catalan Institute for Research and Advanced Studies, 08010 Barcelona, Spain}
\date{\today}

\title{Two-Dimensional Phononic Crystals: Disorder Matters}

\begin{abstract}
The design and fabrication of phononic crystals (PnCs) hold the key to control the propagation of heat and sound at the nanoscale. However, there is a lack of experimental studies addressing the impact of order/disorder on the phononic properties of PnCs. Here, we present a comparative investigation of the influence of disorder on the hypersonic and thermal properties of two-dimensional PnCs. PnCs of ordered and disordered lattices are fabricated of circular holes with equal filling fractions in free-standing Si membranes. Ultrafast pump and probe spectroscopy (asynchronous optical sampling) and Raman thermometry based on a novel two-laser approach are used to study the phononic properties in the gigahertz (GHz) and terahertz (THz) regime, respectively. Finite element method simulations of the phonon dispersion relation and three-dimensional displacement fields furthermore enable the unique identification of the different hypersonic vibrations. The increase of surface roughness and the introduction of short-range disorder are shown to modify the phonon dispersion and phonon coherence in the hypersonic (GHz) range without affecting the room-temperature thermal conductivity. On the basis of these findings, we suggest a criteria for predicting phonon coherence as a function of roughness and disorder.

\end{abstract}


\maketitle

Phononic crystals (PnCs) constitute an attractive class of materials with the potential to manipulate and control the propagation of vibrational energy, i.e., sound and heat. Caused by their periodic structure, these materials can exhibit complete acoustic band gaps due to Bragg reflections and local resonances controlled by geometry and material properties~\cite{Sigalas1992, Sigalas1993, Kushwaha1993, Pennec2010, Still2008, Khelif2010, Maldovan2013}. The periodic modulation of their elastic properties~\cite{Khelif2006, Gorishnyy2007, Still2008, Pennec2010, Schneider2012, Graczykowski2014a, Graczykowski2015} and the reduced dimensionality in nanostructures~\cite{Groenen2008, Cuffe2012, Chavez-Angel2014} lead to strong modifications of the acoustic phonon dispersion which directly affects the phonon group velocity, phonon propagation, and, ultimately, sound and heat transport. 

The continuing miniaturization and progress in nanofabrication techniques have enabled the reduction of the characteristic sizes of PnCs to the nanometer scale and thereby allow the modification and control of phonon propagation and transport properties in the frequency range from hypersonic (GHz)~\cite{Gorishnyy2007, Schneider2012, Graczykowski2014a, Graczykowski2015} to thermal (THz) phonons~\cite{Yu2010, Hopkins2011, Maldovan2013, Maldovan2013a, Zen2014, Neogi2015, Maldovan2015}. The prospect to tailor the thermal conduction and heat capacity has recently triggered tremendous research activities and several authors have reported the successful reduction of the room temperature thermal conductivity in PnCs impacting potential applications in thermoelectricity~\cite{Song2004, Tang2010, Yu2010, Hopkins2011, Reinke2011, Alaie2015, Nakagawa2015, Nomura2015}. The ability to modify the phonon dispersion relation in the hypersonic frequency range, and thus the group velocity of acoustic phonons, has paved the way to applications in RF communication technologies and optomechanics~\cite{OlssonIII2009, Eichenfield2009, Safavi-Naeini2014, Gomis-Bresco2014, Volz2016}. However, studies of the GHz phonon dispersion relation have been performed mostly for bulk~\cite{Gorishnyy2007, Still2008, Schneider2012, Gomopoulos2010, Sato2012, Parsons2014} and surface PnCs~\cite{Graczykowski2014a, Mielcarek2012, Hou2014} including band structure mapping of surface acoustic modes~\cite{Maznev2011, Veres2012} whereas the effects of hole patterning and pillar growth in combination with second order periodicity in thin membranes were only recently studied~\cite{Graczykowski2015}. In particular, the experimental exploitation of disorder, a key aspect in the design of modern photonics~\cite{Wiersma2013} for the guidance and trapping of light and ultrasound by Anderson localization~\cite{Anderson1958, Schwartz2007, Mascheck2012, Hu2008, Yu2009}, is almost completely unexplored in phononic crystals to the present day~\cite{Maire2015}.  

\begin{figure*}[ht]
\begin{center}
\includegraphics[width=\linewidth]{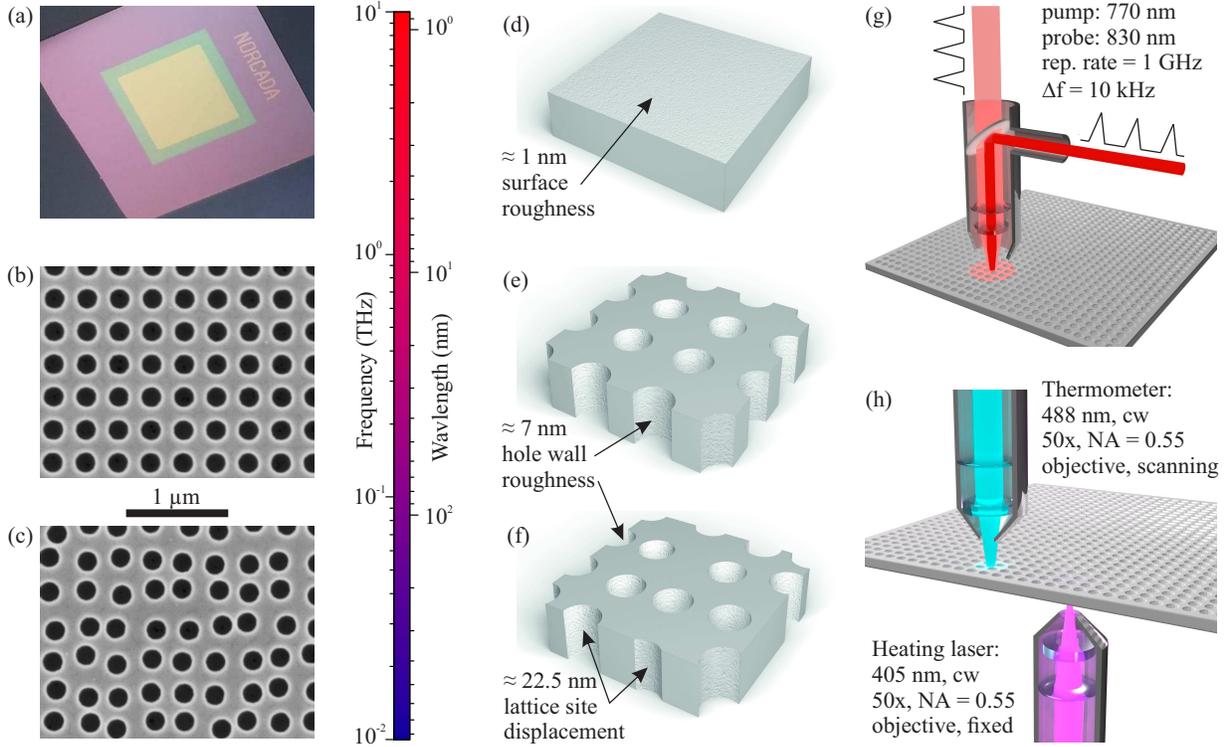}
\caption{(a) Optical image of a 250 nm thick Si membrane, (b) scanning electron microscopy image of Si membrane-based ordered 2D phononic crystal with hole diameter of 175~nm and pitch of 300 nm, (c) disordered PnC with equal hole diameter and filling fraction $\phi=0.267$, (d) schematic of unpatterned membrane with surface roughness, (e) hole wall roughness in ordered PnC, (f) combination of hole wall roughness and lattice site displacement in disordered PnCs, (g) schematic illustration of femtosecond pump-probe reflectivity measurements of an ordered PnC, (h) schematic illustration of 2-laser Raman thermometry measurements of an ordered PnC.}
\label{fig1}
\end{center}
\end{figure*}
 
In this work, we investigate the influence of short-range disorder in Si membrane-based 2D phononic crystals on the GHz and THz phononic properties. We use time-resolved femtosecond pump-probe spectroscopy based on asynchronous optical sampling (ASOPS) to measure the zone-center phonon spectrum and phonon dynamics in the time domain. Finite element method (FEM) simulations are applied to calculate the phonon dispersion relation, 3D displacement fields, and amplitudes of the different mechanical modes. The thermal conductivity of the ordered and disordered PnCs is measured by the recently developed contactless technique of 2-laser Raman thermometry, here applied to PnCs for the first time.

2D phononic crystals were fabricated of free-standing silicon membranes (Norcada Inc.) using electron beam lithography and reactive ion etching to generate ordered and disordered hole patterns with equal filling fractions (Fig.~\ref{fig1}(a-c))~\cite{Sledzinska2016}. The disorder was introduced by random displacements of the holes in $x$ and $y$ direction within the unit cell of the PnC lattice. The hole positions of the disordered PnC were defined by $p=p_0\pm\epsilon\cdot s$, where $p$ is the displaced hole position along the two in-plane axes, $p_0$ is the ordered lattice position, $\epsilon$ is a random number between 0 and 1 and $s$ is the maximum displacement which was set to 45~nm. The level of disorder in percentage of the period $a=$~300~nm is than quantified by $n=s/a\cdot 100\% = 15\%$. Figs.~\ref{fig1}(d-f) display schematic illustrations of the unprocessed membrane with a surface roughness of about 1~nm (d), the ordered PnC with a hole wall roughness of about 7~nm (e), and the disordered PnC with an average displacement of the holes from the ordered lattice sites of 22.5~nm in x- and y-direction (f). 

The coherent acoustic phonon dynamics of the ordered and disordered PnCs with the same filling fractions are investigated using femtosecond pump-probe reflectivity measurements using asynchronous optical sampling~\cite{Bartels2007, Hudert2009, Bruchhausen2011} (see Fig.~\ref{fig1}(g) and section Methods for details). The optical excitation of acoustic vibrations arises from the electronic and thermal stresses induced by the pump pulse which are determined by the generated electron-hole pair density and the temperature-induced lattice deformation, respectively~\cite{Thomsen1986, Wright1995, Hudert2009}. The excited out-of-plane (dilatational) oscillations change the optical cavity thickness of the membrane which leads to a modulation of the probed reflectivity by the Fabry-Perot effect~\cite{Hudert2009, Cuffe2013, Schubert2014}.

\begin{figure*}[ht]
\begin{center}
\includegraphics[width=\linewidth]{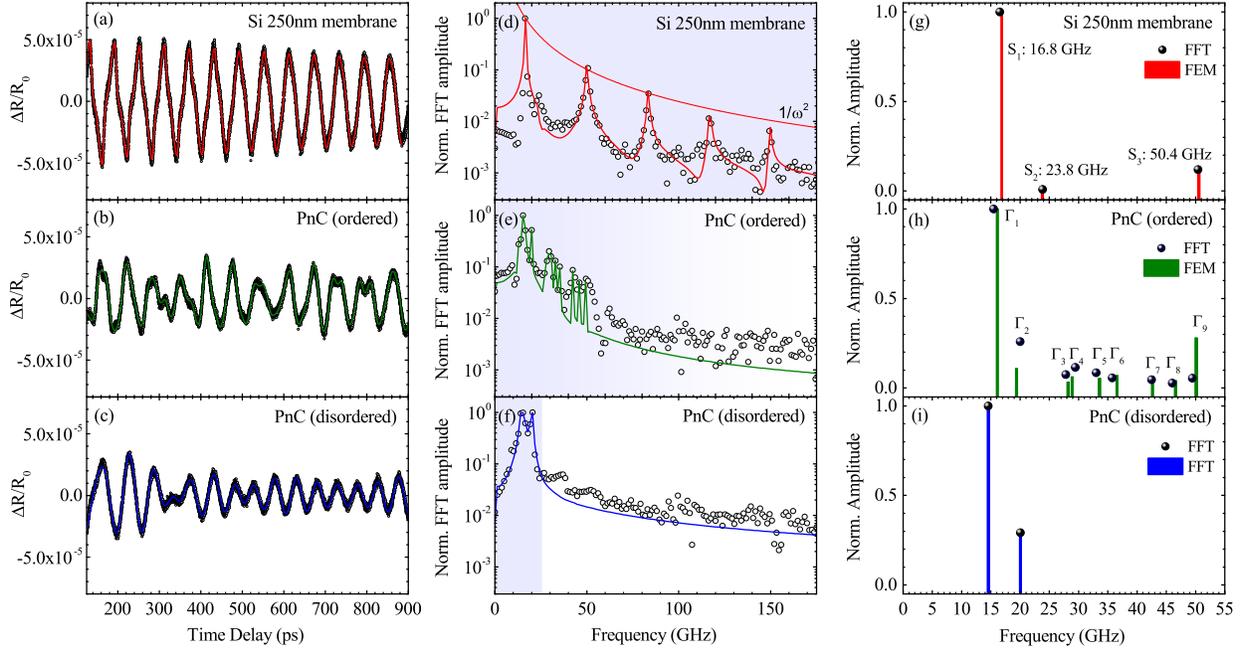}
\caption{Time-resolved reflectivity spectra and corresponding damped sinusoidal fits of (a) unstructured Si membrane, (b) ordered phononic crystals, (c) disordered phononic crystal. Center (d-f): Fourier transformation of measured (dots) and fitted (lines) time-resolved reflectivity spectra. Right (g-i): Normalized amplitude and acoustic phonon frequencies as obtained by FFT of time resolved reflectivity spectra (dots) and finite element method calculations (bars).}
\label{fig2}
\end{center}
\end{figure*}	

Figs.~\ref{fig2}(a-c) show the time-resolved intensity modulation of the reflected probe laser after subtraction of the electronic contribution and system response correction (see SI) for a 250~nm thick Si membrane before patterning (a), with ordered hole lattice (b) and with disordered holes (c). The frequencies of the coherent acoustic phonon modes are derived from the time-domain spectra by numerical Fourier transformation as shown in Fig.~\ref{fig2}(d-f), where dotted spectra represent the FFT spectra of the measured time domain spectra and solid lines are the FFT spectra of the corresponding multi-sinusoidal fits in Figs.~\ref{fig2}(a-c). The blue shaded areas indicate the frequency ranges in which coherent acoustic phonons are detected. It should be noted that only phonon modes with amplitudes greater than the noise level are displayed in the fits, thus, the highest phonon frequency is not a strict limit for phonon coherence as indicated by the gradient in the blue shaded range. Using this experimental approach, we can directly obtain the complete zone-center ($q=$~0) coherent phonon spectrum from the GHz to the THz regime. In the case of the bare membrane (Fig.~\ref{fig2}(d)), the different harmonics in the vibrational spectrum appear as equidistant peaks as a consequence of the confinement of the acoustic modes~\cite{Hudert2009, Torres2004, Groenen2008}. The lowest frequency peak at 16.8~GHz thereby corresponds to the first order symmetric ($S_1$) mode and higher frequency modes up to the 9th harmonic at 151 GHz are clearly visible. The absence of even harmonics can be understood taking into account that those modes have only in-plane displacement at the $\Gamma$ point so that no modulation of the optical cavity thickness occurs~\cite{Schubert2014}. Using the value of the longitudinal sound velocity for Si [001] of $v_L=8433$~m/s~\cite{McSkimin1964}, the thickness of the membrane $d$ determines the frequencies of the observed modes $f_n = nv_L/2d$, where $f_n$ is the frequency of the $n$-th harmonic ($n=1,3,5,...$) of the symmetric (dilatational) mode~\cite{Hudert2009}. In addition, a weak signal at 23.8~GHz is observed (0.03 of the amplitude of $S_1$). Using FEM modeling, we identify this peak as the symmetric $S_2$ mode which might be visible due to excitation of propagating in-plane modes with small but nonzero wave numbers and therefore nonzero out-of-plane displacement~\cite{Auld1990}.

Following the discussion of the acoustic phonon dynamics of the non-patterned membrane, we now focus on the modification of the frequency spectrum in ordered and disordered PnCs by hole patterning of the original membrane. The time resolved pump-probe reflectivity spectra for the ordered and disordered PnCs are displayed in Fig.~\ref{fig2}(b) and \ref{fig2}(c), respectively. The periodic signal of the unpatterned membrane in Fig.~\ref{fig2}(a) is replaced by a more complex time response of the reflectivity change in the ordered and disordered PnCs. This indicates a strong modification of the phonon dispersion relation with the appearance of additional acoustic phonon modes that contribute to reflectivity modulations. The corresponding frequency spectra for the ordered and disordered PnCs are shown in Fig.~\ref{fig2}(e) and \ref{fig2}(f). The relative amplitudes of the different modes, normalized to the most intense mode at about 16~GHz, are displayed in Fig.~\ref{fig2}(g-i) (dots) and compared to the results of FEM simulations (bars).

In order to explain the observed differences between the unpatterned membrane and the ordered PnC, we calculate the acoustic phonon dispersion relation and the 3-dimensional displacement fields for the zone-center modes up to 55~GHz by means of FEM modeling as described in Ref.~\cite{Graczykowski2015}. In principle, the reflectivity of the membrane is modulated by the induced mechanical modes which result in non-zero average thickness variations. Here the model assumes a predominant role of the optical cavity thickness mechanism with negligible contribution of the photoelastic effect (PE)~\cite{Hudert2009}. The pump-induced variation of the membrane thickness $\Delta d$ is of the order of a picometer and directly proportional to the measured relative change of the reflectivity $\Delta R/R_0$. In the case of the PnCs the mechanical modes are complex and $\Delta d$ is position dependent. We correlate the amplitudes $A_i$ of the modes $\omega_i$ with their corresponding average change of thickness $|\overline\Delta d|$ calculated over the whole FEM unit cell using the formula: 
      
\begin{equation}
\label{eq1}
A_i(\omega_i)\propto\overline{\Delta d}\propto\frac{1}{S\omega_i}\int_S \big((u_i(z=0)-u_i(z=d)\big)\mathrm{d}S,
\end{equation}

\noindent where $u_i(z)$ are the out-of-plane displacement components, and $S$ is the free surface area. The displacement fields of all the FEM solutions are normalized in such a way that all the modes store the same elastic energy and are populated according to the Planck distribution at high temperature.

\begin{figure*}[t!]
\begin{center}
\includegraphics[width=\linewidth]{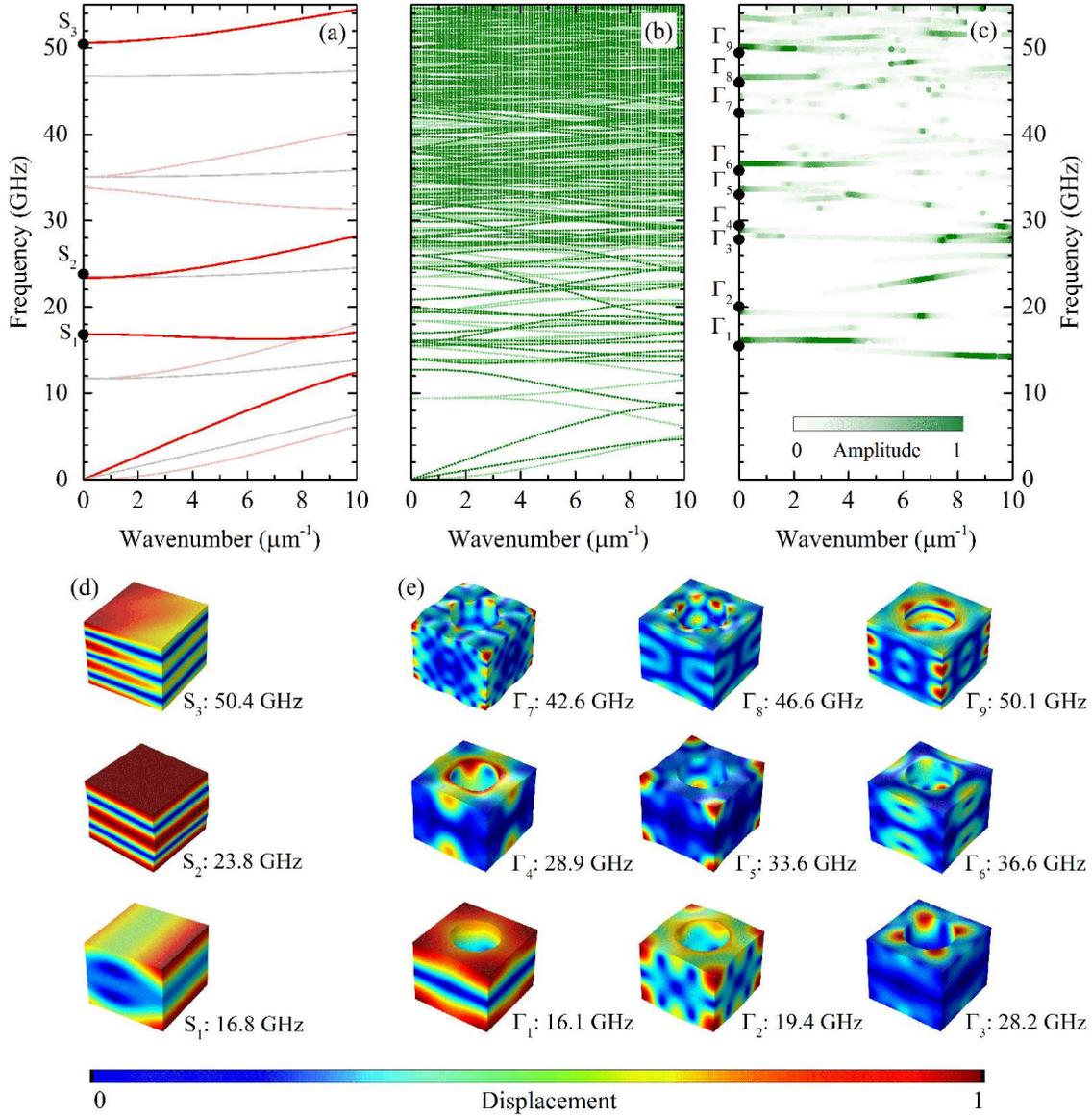}
\caption{Calculated acoustic phonon dispersion relation for a 250~nm thick Si membrane without pattern in [110] direction (a) and with ordered hole pattern in $\Gamma X$ direction (2D PnC) (b) showing symmetric (dark color), antisymmetric (light color) and shear horizontal (gray) modes. Dispersion relation displaying the amplitude of the out-of-plane modes in ordered PnCs (c). Black dots indicate the experimentally obtained zone-center phonon frequencies (b) and (c). The corresponding 3D displacements fields for the symmetric modes at the $\Gamma$ point are shown for the unpatterned membrane (d) and the ordered PnC (e). Only modes with non-vanishing out-of-plane displacement are displayed.}
\label{fig3}
\end{center}
\end{figure*}

The acoustic phonon dispersion relations for the membranes before and after hole patterning are displayed in Fig.~\ref{fig3}(a) and \ref{fig3}(b), respectively. For the unpatterned membrane, the first three symmetric modes $S_1$, $S_2$, and $S_3$ are precisely reproduced by the FEM simulations regarding both amplitude and frequency (Fig.~\ref{fig2}(g) and Fig.~\ref{fig3}(a)). The decreasing amplitude of the higher harmonics can be accurately described by Eq.~(\ref{eq1}) which in the case of the unpatterned membranes simplifies to a $1/\omega^2$ relation (Fig.~\ref{fig2}(d))~\cite{Hudert2009}. The fact that the mode amplitudes obey this relation indicates that the phonon coherence is not destroyed by, e.g., surface roughness up to at least 150~GHz. In the case of the ordered PnCs, the strong modifications of the phonon dispersion relation in Fig.~\ref{fig3}(b) compared to the bare membrane (Fig.~\ref{fig3}(a)) arises from band folding and band splitting when the Bragg condition is satisfied~\cite{Graczykowski2015}. Considering the multitude of different modes in the dispersion relation, the question arises why there are only nine discrete modes visible in the time domain measurements. The answer becomes clear, if we consider the out-of-plane displacement amplitudes of the different modes according to Eq.~\ref{eq1} which are plotted together with the measured zone-center phonon frequencies in Fig.~\ref{fig3}(c). In excellent agreement with the experimental results, the calculations reveal nine discrete modes in the frequency range up to 55~GHz ($\Gamma_1 - \Gamma_9$) with non-vanishing out-of-plane displacement amplitude at the $\Gamma$ point. The corresponding 3-dimensional displacement fields for these modes are displayed in Fig.~\ref{fig3}(d) and \ref{fig3}(e) enabling an unambiguous identification of the displacement characteristic of each observed modes. 

Comparing the acoustic phonon spectrum of the ordered PnCs with those of the disordered PnCs, it is apparent that the dispersion relation and zone-center phonon frequencies differ significantly. In a disordered PnC, no individual modes above about 20~GHz can be detected (see Figs.~\ref{fig2}(f) and \ref{fig2}(i)). This observation demonstrates the importance of translational symmetry to build coherent phonon modes in the hypersonic frequency range. Interestingly, two modes remain observable which are comparable in frequency and amplitude to the lowest frequency modes $\Gamma_1$ and $\Gamma_2$ of the ordered PnC, although significantly broadened. The fact that these modes are largely unaffected by disorder can be understood considering the 3D displacement fields of the lowest frequency modes in the unpatterned membrane ($S_1$) and the ordered PnC ($\Gamma_1$) in Figs.~\ref{fig3}(d) and \ref{fig3}(e), respectively. Both modes exhibit a large out-of-plane displacement amplitude and similar displacement symmetry indicating that they are mainly governed by the bare membrane and do not depend on the second-order periodicity. Consequently, these modes are not significantly affected by the degree of order / disorder in the PnCs. Finally, it should be noted that the frequency of the lowest energy mode in the disordered PnC (14.6~GHz) and ordered PnC (15.5~GHz) is slightly smaller compared to the reference membrane (16.8~GHz) which is caused by the softening of the material due to the combined effects of mass removal and periodic arrangement. 

\begin{figure*}[ht]
\begin{center}
\includegraphics[width=\linewidth]{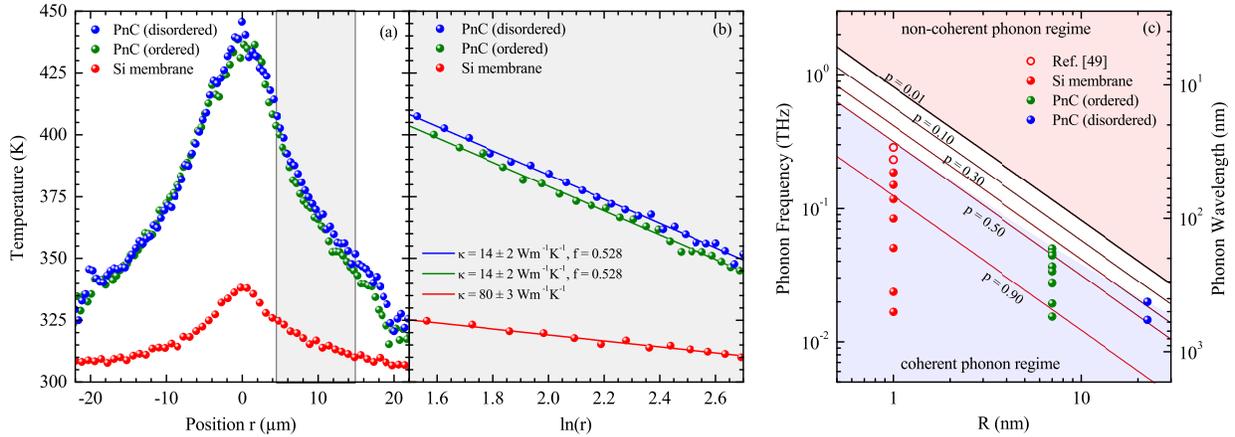}
\caption{(a) Temperature line-scan profiles of the ordered and disordered PnCs and the unpatterned Si membrane for a localized thermal excitation using two-laser Raman thermometry (2LRT). (b) Logarithmic plot of the highlighted area in (a) to visualize the ln($r$) relation as predicted by Eq.~\ref{eq2}. (c) Phonon frequency / phonon wavelength as function of characteristic size $R$ (surface roughness, hole wall roughness, lattice site displacement) for selected specularity parameters $p$ between 0.01 and 0.90. Data points represent measured coherent acoustic phonon frequencies. The blue shaded area indicates the coherent phonon regime extrapolated from the highest measured coherent phonon frequencies, the red shaded area marks the non-coherent phonon regime.}
\label{fig4}
\end{center}
\end{figure*}

Up to this point, we have limited our discussion to the hypersonic (GHz) frequency range of the phonon spectrum. Taking into account that no coherent phonon modes in the ordered and disordered PnCs could be observed at frequencies above 55 and 20~GHz, respectively, we use a different approach to investigate the influence of order and disorder on the thermal properties: Two laser Raman thermometry (2LRT)~\cite{Reparaz2014}. The main advantage of this technique with respect to, for example, electrical measurements or time-domain thermoreflectance (TDTR), is given by its contactless nature avoiding the introduction of additional thermal interface resistances. Thus, the thermal conductivity of the PnCs can be directly obtained from the measurements without additional modeling. A spatially fixed heating laser generates a localized steady-state thermal excitation, whereas a low power probe laser measures the spatially-resolved temperature profile with sub-micrometer resolution through the temperature dependent Raman frequency of the optical phonons in the material. Fig.~\ref{fig4}(a) displays the temperature profiles for ordered and disordered PnCs and the unpatterned Si membrane obtained by 2LRT with the experimental arrangement shown schematically in Fig.~\ref{fig1}(h). Applying Fourier's law in 2-dimensions for a thermally isotropic medium leads to a temperature field $T(r)$ with:

\begin{equation}
\label{eq2}
 T(r)=T_0+\frac{P_{abs}}{2\pi f d \kappa_0}ln(r/r_0)
\end{equation}

\noindent where $(r_0,T_0)$ is an arbitrary point in the temperature field, $P_{abs}$ is the absorbed power, $f=0.528$ is a correction factor for the missing material due to the holes in the PnCs with a filling fraction of $\phi=0.267$~\cite{Alaie2015}, $d$ is the thickness of the PnC membranes, and $\kappa_0$ is the thermal conductivity. Here, $\kappa_0$ can be treated as temperature independent since the temperature range is sufficiently small ($\approx$ 50 K). We recall that for the case of bulk Si the thermal conductivity changes by about 15\% in the range from 350~K to 400~K \cite{Glassbrenner1964}. This variation represents only an upper (bulk) limit since the temperature dependence is typically reduced as boundary scattering increases. Fig.~\ref{fig4}(b) displays the thermal decays in logarithmic scale according to Eq.~(\ref{eq2}), thus, the slope of the thermal decay is directly related to $\kappa_0$. The purely linear decay observed in this graph validates the temperature independent treatment of $\kappa$. A deviation from this linear relation is expected in cases where $\kappa=\kappa(T)$ as discussed in Ref.~\cite{Reparaz2014}. Based on these measurements, we obtain the same value for the thermal conductivity $\kappa_0=14\pm2$ Wm$^{-1}\mathrm{K}^{-1}$ for the ordered and disordered PnCs compared to $\kappa_0=80\pm3$ Wm$^{-1}\mathrm{K}^{-1}$ in case of the unpatterned membrane. Considering the correction factor for the material loss in the holes of the PnCs, we would obtain an effective thermal conductivity of the PnCs in the absence of size effects of $f\kappa_0=46\pm3$ Wm$^{-1}\mathrm{K}^{-1}$.

The reduction in the thermal conductivity of the PnCs down to $18$\% of the value of the unpatterned membrane, about a six-fold reduction, is in accordance with other recent studies of the thermal conductivity of PnCs with comparable dimensions~\cite{Tang2010, Yu2010, Zen2014, Alaie2015, Nakagawa2015, Nomura2015}. This drastic reduction cannot be solely explained by the mass loss due to hole patterning. Instead, two aspects need to be considered: (i) diffuse boundary scattering and (ii) phonon coherence. A decrease of thermal conductivity by diffuse boundary scattering is expected due to the increase of the surface area caused by the introduction of holes as well as the hole wall roughness of about 7~nm due to the patterning process (c.f. Figs.~\ref{fig1}(e) and \ref{fig1}(f)). We address the issue of phonon coherence by plotting in Fig.~\ref{fig4}(c) the phonon frequencies of the measured coherent acoustic phonons (Fig.~\ref{fig2}) as function of the characteristic size $R$ limiting in each case the measured phonon frequency range, i.e. surface roughness, hole wall roughness, and average lattice site displacement. Next, we plot the corresponding phonon frequencies for selected specularity parameters $p$ as function of $R$:  
\begin{equation}
f(R)|_{p}=\sqrt{\frac{-\mathrm{ln}(p)}{16\pi^3}}\cdot\frac{v_L}{R}
\end{equation}
where the specularity parameter $p$ expresses a quantitative measure for the percentage of specular scattering at a normal surface with given roughness (1 for purely specular scattering and 0 for purely diffusive scattering). The dependence of $p$ as function of the phonon wavelength is displayed for given values of $R$ in SI Fig.~5. Attempting to derive a general criteria for the non-coherent phonon regime (fully diffusive phonon scattering), we extrapolate the specularity parameter towards 0 in SI Fig.~5. The corresponding phonon wavelength is given in good approximation by the line for $p=$~0.01 in Fig.~\ref{fig4}(c). By plotting the phonon wavelength $\lambda_{ph}$ as function of roughness for this specularity parameter, we obtain $\lambda_{ph}\leq10R$ as simple criteria for the non-coherent phonon regime. Following this approach, we obtain frequency limits of 800~GHz, 115~GHz, and 36~GHz for surface roughness values of 1~nm, 7~nm, and 22.5~nm, respectively. It is important to note that the specularity parameter as introduced by Ziman~\cite{Ziman1962} only considers the surface roughness for a normal incidence wave, not wall roughness or lattice site displacement. However, despite the different types of roughness in our phononic crystals, the computed values reproduce the general tendency of the measured decreasing high frequency limit of coherent phonons in our phononic crystals. In fact, our experimental data suggests that phonon coherence is already affected by roughness corresponding to a specularity parameter between 0.3 and 0.5 (c.f. Fig.~\ref{fig4}(c)). Using the more conservative value of $p=$~0.5, we find in a rough approximation that $\lambda_{ph}>25R$ constitutes a realistic criteria for the coherent phonon regime. Consequently, we suggest that disorder, quantified by the average hole displacement from the periodic lattice sites, can also be considered as a type of roughness for long wavelength phonons as can be seen when plotting the measured coherent phonon frequencies of the disordered PnCs for a roughness $R=25$~nm in Fig.~\ref{fig4}(c): the measured frequencies are in good agreement with the calculated frequency for $p=$~0.3 - 0.5. The absence of any coherent acoustic phonon signal with $f>20$~GHz for the disordered PnCs in Fig.~\ref{fig2}(f) and \ref{fig2}(i) can then be understood as approaching the frequency limit of the coherent phonon regime for this given level of disorder.

It now becomes clear why no differences between the thermal conductivity of the ordered and disordered PnCs should be expected due to coherent effects as confirmed by the equal values of $\kappa$ measured in the 2LRT experiments: On the one hand, the hole wall roughness of the ordered and disordered PnCs of about 7~nm prevents coherent effects for phonons with frequencies above $\approx$100~GHz (see Fig.~\ref{fig4}(c)) whereas on the other hand, the frequencies of the dominant thermal phonons are in the low THz regime. In other words, the wavelengths of thermal phonons at room temperature are below 10~nm and therefore commensurate with the characteristic roughness $R$ in the studied samples which excludes any coherent effects based on the above given criteria. Consequently, the thermal conductivity is not affected by phonon coherence for both, the ordered and the disordered PnCs. Even for a roughness value as low as 1~nm, we predict phonon coherence only up to about 400~GHz at room temperature. According to Fig.~\ref{fig4}(c) modifications of the thermal conductivity due to phonon coherence will only occur for very smooth surfaces/interfaces where the limit of the coherent phonon regime reaches the THz range or for very low temperatures where the wavelength of the thermal phonons is significantly enlarged and thereby greatly exceeds the characteristic roughness $R$. These results are also in agreement with a recent study of Maire et. al~\cite{Maire2015}, who observed pronounced modifications of the thermal conductivity depending on the level of disorder in Si PnCs only for temperatures up to about 10~K.

In conclusion, we have addressed the question to what extend disorder influences the phononic properties of 2-dimensional phononic crystals both in the GHz and THz frequency range. In a first step, we have shown that patterning of a 2D Si membrane by an ordered phononic crystal lattice strongly modifies the frequencies and dispersion relation of hypersonic vibrations measured by femtosecond time-domain spectroscopy. Using finite element method simulations we have uniquely identified the displacement characteristic of each mode by the calculation of the dispersion relation and the 3-dimensional displacement fields. In particular, we have developed a simple model that accurately predicts the amplitudes of the out-of-plane displacement for all observed modes in the ordered PnCs. The introduction of disorder in the PnCs drastically modifies the hypersonic phonon spectrum resulting in the suppression of coherent acoustic phonon modes. Measurements of the thermal conductivity using a novel two laser Raman thermometry technique have shown that a six-fold reduction of the thermal conductivity occurs for both ordered and disordered PnCs ($\kappa_0=14\pm2$ Wm$^{-1}\mathrm{K}^{-1}$) with respect to the unpatterned membrane ($\kappa_0=80\pm3$ Wm$^{-1}\mathrm{K}^{-1}$). Based on the measured coherent acoustic phonon frequencies for different levels of roughness and disorder we have derived two criteria for the prediction of coherent and non-coherent phonon regimes: (i) phonon coherence is unaffected if the roughness $R$ is smaller than 1/25 of the phonon wavelength and (ii) phonon coherence is destroyed if $R$ is greater than 1/10 of the phonon wavelength. The results reveal the impact of surface roughness and disorder on the observation of coherent effects in phononic crystals and demonstrate that the room temperature thermal conductivity in comparable phononic crystals should not be affected by the change of the phonon dispersion resulting from coherent boundary scattering.

\section{Materials and Methods}

\noindent\textbf{Sample preparation:} Commercially available single crystalline silicon (100) membranes (Norcada Inc.) with a thickness of 250 nm and window size of 3.2$\times$3.2~mm$^2$ were used to fabricate 2D PnCs~\cite{Sledzinska2016}. PMMA 950k (Allresist) was spun at 4000 rpm for one minute, followed by a 60 min bake at 100$^\circ$C in an oven. Electron beam lithography (Raith 150-TWO) was carried out to pattern ordered and disordered PnCs with a hole size of 175 nm and a pitch of 300 nm for the ordered PnCs and equal filling fraction of $\phi=0.267$ for the disordered PnCs (c.f. Fig.~\ref{fig1}). The dimensions of the structures were $\mathrm{50\times50}$ $\mathrm{\mu}$m. After development in 1:3 methyl isobutyl ketone:isopropanol (MIBK:IPA), the samples were post-baked for 1 min in 80$^\circ$C on a hot plate. The pattern was transferred to silicon using the reactive ion etching Bosch process (Alcatel AMS-110DE) and finally the samples were cleaned in an oxygen plasma system (PVA Tepla).

\noindent\textbf{Femtosecond pump-probe reflectivity measurements:} Femtosecond pump-probe reflectivity measurements based on the asynchronous optical sampling (ASOPS) technique~\cite{Bartels2007, Hudert2009, Bruchhausen2011} were used to investigate the acoustic phonon dynamics of ordered and disordered PnCs and unpatterned membranes in the hypersonic (GHz) frequency range. The experimental method is based on two asynchronously coupled Ti:sapphire ring cavity lasers with a repetition rate of 1 GHz and a nominal pulse length of about 50 fs. The time delay between the pump and probe pulses is achieved through an actively stabilized frequency offset of 10 kHz between the repetition rate of the two laser oscillators. This allows for a linearly increasing time delay between pump and probe pulses with steps of 10 fs without the need for a mechanical delay line. Owing to the high repetition rate of 1 GHz, an excellent signal-to-noise ratio of above $10^7$ can be achieved for typical acquisition times in the seconds to minutes range. Pump-probe reflectivity measurements were performed by focusing both lasers collinear onto the PnC membranes using a 50$\times$ microscopy objective (Olympus, NA = 0.55) in normal incidence geometry, resulting in a spot size of about 2~$\mu$m which corresponds to an excitation area in the PnC of about 35 unit cells. The pump laser was tuned to a center wavelength of 770 nm with an average power of 8 mW and the probe laser to 830 nm with a power of 4 mW. The reflected probe laser was spectrally filtered by a long-pass filter at 800 nm to eliminate contributions from the pump laser and recorded with a low noise photodetector with 125 MHz bandwidth.

\noindent\textbf{Two-laser Raman thermometry:} Thermal conductivity measurements were conducted using two-laser Raman thermometry, a novel technique recently developed to investigate the thermal properties of suspended membranes ~\cite{Reparaz2014}. A spatially fixed heating laser generates a localized steady-state thermal excitation, whereas a low power probe laser measures the spatially-resolved temperature profile with sub-micrometer resolution through the temperature dependent Raman frequency of the optical phonons in the material. Both lasers were focused on the PnCs using 50$\times$ microscope objectives with numerical apertures of NA = 0.55. The power of the heating laser with a wavelength of $\lambda_{heat}=\mathrm 405$~nm was set to 1~mW and the power of the probe laser with a wavelength of $\lambda_{probe}=\mathrm 488$~nm to 0.1~mW in order to avoid local heating by the probe laser while measuring the temperature field. The absorbed power is measured for each sample as the difference between incident and transmitted plus reflected light intensities probed by a calibrated system based on a non-polarizing cube beam splitter. The measurements were performed at ambient pressure which introduces heat losses through convective cooling. This effect accounts for about 30\% of the thermal conductivity in our samples, i.e., the measured values for the thermal conductivity of the PnCs were $\kappa_0=21\pm2$ Wm$^{-1}\mathrm{K}^{-1}$. After correcting for the heat transport due to convective cooling we obtained the reported value of $\kappa_0=14\pm2$ Wm$^{-1}\mathrm{K}^{-1}$ for the ordered and disordered PnCs. A detailed discussion of the influence of convective cooling on the experimental values obtained for the thermal conductivity in Si PnCs is being published elsewhere~\cite{Graczykowski2016}.

\section{ASSOCIATED CONTENT}
\textbf{Supporting Information}

The Supporting Information is available free of charge on the ACS Publications website at DOI:

Description of the finite element modeling simulations; femtosecond pump-probe reflectivity spectra for a bare membrane, an ordered PnC, and a disordered PnC; specularity parameter as function of phonon wavelength for selected roughness values.

\section{AUTHOR INFORMATION}

\textbf{Corresponding Author}
Email: wagner.markus.r@gmail.com

\noindent\textbf{Present Addresses}
M.R.W.: TU Berlin, Institute of solid state physics, Hardenbergstr. 36, 10623 Berlin, Germany

\section{ACKNOWLEDGMENTS}
The authors acknowledge financial support from the EU FP7 project MERGING (Grant No. 309150), NANO-RF (Grant No. 318352) and QUANTIHEAT (Grant No. 604668); the Spanish MICINN projects nanoTHERM (Grant No. CSD2010-0044) and TAPHOR (Grant No. MAT2012-31392); and the Severo Ochoa Program (MINECO, Grant SEV-2013-0295). M.R.W. acknowledges the postdoctoral Marie Curie Fellowship (IEF) HeatProNano (Grant No. 628197).

\end{document}